\documentclass[ reprint, amssymb, amsmath, aip,cha]{revtex4-1}

\usepackage[colorlinks=true,linkcolor=blue]{hyperref}%
\usepackage{graphicx}
\usepackage{amsmath}
\newcommand{\vc}{\mathbf}

\begin{document}

\title{First experiments on Revolver shell collisions at the OMEGA Laser}%

\author{Brett Scheiner}
\email{bss@lanl.gov}
\thanks{Also at the Department of Physics and Astronomy, University of Iowa}
\affiliation{Los Alamos National Laboratory}
\author{Mark J. Schmitt}
\affiliation{Los Alamos National Laboratory}
\author{Scott C. Hsu}
\affiliation{Los Alamos National Laboratory}
\author{Derek Schmidt}
\affiliation{Los Alamos National Laboratory}
\author{Jason Mance}
\affiliation{Mission Support and Test Services}
\author{Carl Wilde}
\affiliation{Los Alamos National Laboratory}
\author{Danae N. Polsin}
\affiliation{Laboratory for Laser Energetics, University of Rochester}
\author{Thomas R. Boehly}
\affiliation{Laboratory for Laser Energetics, University of Rochester}
\author{Frederic J. Marshall}
\affiliation{Laboratory for Laser Energetics, University of Rochester}
\author{Natalia Krasheninnikova}
\affiliation{Los Alamos National Laboratory}
\author{Kim Molvig}
\affiliation{Los Alamos National Laboratory}
\affiliation{Massachusetts Institute of Technology}
\author{Haibo Huang }
\affiliation{General Atomics}

\begin{abstract}
Results of recent experiments on the OMEGA Laser are presented, demonstrating the ablator-driver shell collision relevant to the outer two shells of the Revolver triple-shell inertial-confinement-fusion concept  [K. Molvig et al., PRL~{\bf 116}, 255003 (2016)]. These nested two-shell experiments measured the pre- and post-collision outer-surface trajectory of the 7.19 g/cc chromium inner shell. Measurements of the shell trajectory are in excellent agreement with simulations; the measured outer-surface velocity was $7.52\pm0.59$ cm/$\mu$s compared to the simulated value of 7.27 cm/$\mu$s. Agreement between the measurements and simulations provides confidence in our ability to model collisions with features which have not been validated previously.
Notable features include the absence of $\sim$40 mg/cc foam between shells commonly used in double shell experiments, a dense (7.19 g/cc) inner shell representative of the densities to be used at full scale, approximately mass matched ablator payload and inner shells, and the inclusion of a tamping-layer-like cushion layer for the express purpose of reducing the transfer of high mode growth to the driver shell and mediation of the shell collision. Agreement of experimental measurements with models improves our confidence in the models used to design the Revolver ignition target.
\end{abstract}

\maketitle

\section{Introduction}
In conventional hot-spot ignition, the central deuterium-tritium (DT) gas is compressed by an accelerated DT ice layer and ablator driven either by x-ray radiation from a hohlraum or directly by lasers incident on the ablator\cite{2004PhPl...11..339L,2015PhPl...22k0501C}. The compressed low density fuel forms a hotspot from which a thermonuclear burn wave propagates outwards, consuming the high density DT ice layer and resulting in high yield. This ignition scheme, which has dominated inertial confinement fusion (ICF) for decades, has yet to produce ignition in the laboratory, primarily due to the effects of asymmetry and mix\cite{2016PhPl...23e6302C,2013PhRvL.111h5004M}. Ignition schemes that abandon the propagating burn wave concept and instead aim to ignite the fuel volume simultaneously have recently gained interest, with Revolver\cite{2016PhRvL.116y5003M} and Double Shell targets\cite{2018PhPl...25i2706M,DSEXP} being the prime examples.

Revolver is a direct-drive triple-shell ICF concept that aims to turn a 6 ns laser drive pulse on the ablator into a 1.5 ns pressure pulse delivered to the liquid DT fuel by the inner most high-Z (Au or W) pusher shell. The high Z pusher absorbs all of the non-thermal bremsstrahlung radiation from the fuel in a narrow layer of heated metal. This layer emits thermal radiation back into the optically thin fuel cavity, filling it with wall temperature Planckian radiation that contains very little energy. Fuel losses are reduced to the energy flowing to the wall boundary layer. The result is a lower ignition temperature of $T_i\approx2.5$ keV\cite{2016PhRvL.116y5003M,2018PhPl...25i2706M}. 


During the implosion, the fuel is first shock heated to a temperature which scales like implosion velocity squared. Shock collapse marks the end of  the shock heating phase. The shell convergence is, $C \approx 2.2$, independent of the implosion velocity. The adiabatic compression phase that follows is designed to reach ignition at a total convergence C=10. The implosion velocity is determined by the temperature needed out of the shock phase to do this. With adiabatic compression increasing the temperature by convergence squared, $C^2 \approx 20$, and the remaining convergence allotted to this phase of $C_{adiabatic} \approx4.5$, we need a  shocked fuel temperature of around $T\approx150$ eV. 

The electrons and ions are heated in equilibrium with relatively flat radial profiles having only a factor of two decrease in temperature from the fuel center to the pusher wall\cite{2018PhPl...25h2708M}. Unlike hot-spot ignition, there is no propagation of a burn wave into a low temperature fuel. In typical hot spot designs, the PdV work on the fuel must overcome the radiative bremsstrahlung loss, setting a high minimum implosion velocity. By effectively eliminating radiative losses in the Revolver target, one is left with the implosion velocity as a parameter free to serve the purpose as described above and resulting in a simpler, less challenging hydrodynamic scheme. In fact, higher implosion velocity is not necessarily a benefit. Although faster implosions could reduce the convergence required to reach ignition, the fuel would be less compressed and susceptible to premature disassembly\cite{2016PhRvL.116y5003M,2018PhPl...25i2706M}. For the Revolver design, a pusher implosion velocity of $u_I=19$ cm/$\mu$s results in a required pusher convergence ratio of 9.4. 

\begin{figure}
\includegraphics[width=.45\textwidth]{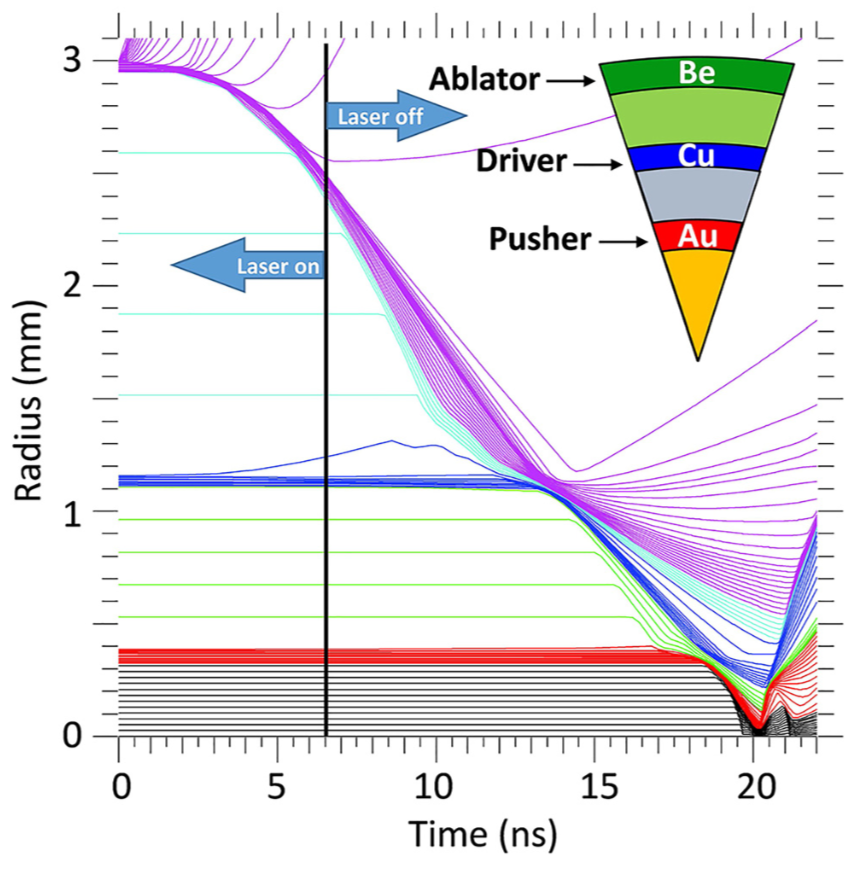}
\caption{A Lagrangian plot of the ignition-scale Revolver implosion including a cutaway pie diagram of the shell configuration. Reproduced with permission from Ref.~\onlinecite{2016PhRvL.116y5003M}. }
\end{figure}

The required implosion velocity of the innermost pusher shell is provided by the middle driver shell, nominally made of Cu, Cr, or some other material in the density range of $\sim$7-10 g/cc. The driver shell is accelerated via a collision with the low density $\sim$1-2 g/cc (Be or CH) ablator at a convergence of 3, and itself collides with the pusher, also at a convergence of 3, providing the energy needed for compression of the fuel. Achieving the required ignition temperature of the DT fuel with low pusher convergence depends on the collision efficiency between the ablator and driver as well as the driver and pusher.

In this paper, we evaluate the predictive capability of computational models of the collision between the ablator and inner shell relevant to a sub-ignition scale two-shell implosion at the OMEGA Laser. Although direct drive double-shell targets have been shot on OMEGA and the SG-II laser facilities previously\cite{2006PhPl...13e6306K,2019NucFu..59d6012T}, the targets utilized in the present experiments have several unique characteristics intrinsic to the Revolver design that have yet to be used in direct drive experiments.
These features are (i) a thin ablator designed for optimal performance in direct drive, (ii) the absence of $\sim$40 mg/cc foam between shells, (iii) a dense (7.19 g/cc) inner shell representative of the densities to be used at full scale, (iv) approximately mass matched ablator payload and driver shells, and (v) the inclusion of a tamping-layer-like cushion layer for the express purpose of reducing the transfer of high mode growth to the inner shell and to mediate the shell collision. With these features in place, we demonstrate the predictability of the shell collision efficiency via simulated and measured velocities of the outer surface of the chromium driver shell. The validation of simulations of collisions with these features provide confidence that the ignition-scale model accurately captures the new features of the collision process which have yet to be tested experimentally until the present work.

These experiments are part of an ongoing effort to evaluate the viability of the Revolver ignition concept. Previous experiments on OMEGA have demonstrated the ability to achieve low levels of laser plasma instabilities and an ablator hydrodynamic efficiency of $\sim10\%$\cite{REXP}.  

This paper is organized as follows, Sec. II gives an overview of the motivation for target design choices and the details of the experiment setup. Sec. III presents the analysis of radiography results and makes a comparison with simulations to evaluate their predictive capability. Concluding remarks are made in Sec. IV and details of the analysis are given in the Appendix.

\section{Experiment Design }

\subsection{Target Design Considerations}

The target design originated from a scaled down version of the outer two shells of the Revolver baseline ignition design (See Fig.~1). In the baseline design, the ablator shell collides with the middle driver shell after a convergence of 3. The convergence of the ablator in these sub-scale experiments is the same in the baseline design. The outer Be shell diameter was set at 1200 $\mu$m with a thickness of $15\mu$m, which set the Cr driver shell outer diameter to 400$\mu$m. The driver shell thickness in the experiments was set so that it would be approximately mass matched to the ablator payload, which is predicted to maximize the momentum transfer between shells. The shell materials were chosen to be consistent with those in the full scale design. Previous double-shell experiments that measured post-collision inner-shell trajectories have opted to substitute a low density inner shell for the high-Z high-density shell material to ease the requirements on diagnostic imaging\cite{2006PhPl...13e6306K,DSEXP}. Such implosions differ from their counterpart ignition designs in the Atwood number of interfaces and in the material properties that determine the evolution of the experiment.  

\begin{figure}
\includegraphics[width=.5\textwidth]{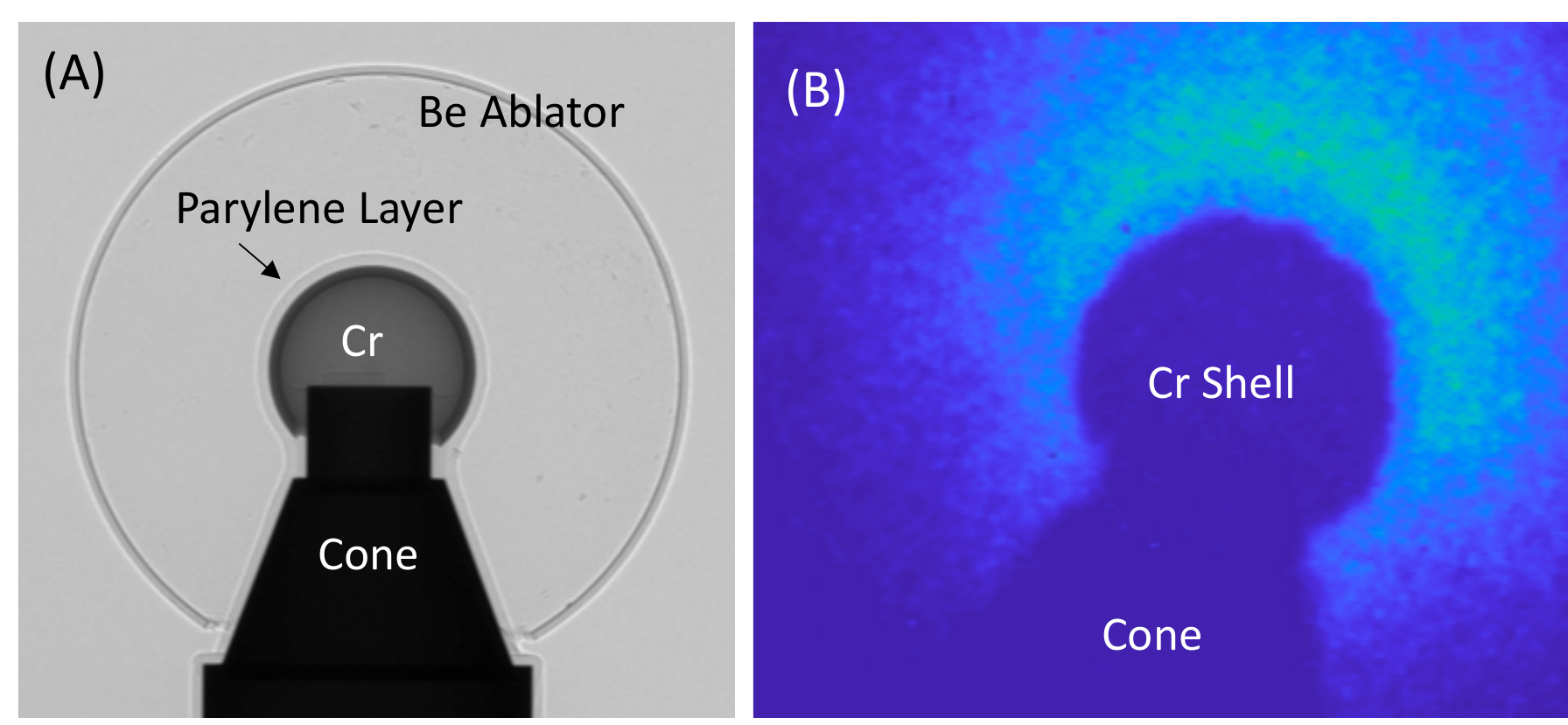}
\caption{\label{radiograph} (A) A pre-shot radiograph showing the Be ablator, Cr inner shell with parylene layer, and gold support cone. (B) A backlit image of the imploding Cr shell and cone.}
\end{figure}

Revolver targets are unique among other multi-shell designs in the absence of a $\sim 40-70$mg/cc foam supporting the inner shell. The Revolver ignition design is unable to tolerate the energy loss incurred by compressing $\sim100 \ \textrm{mm}^3$ of foam (a consequence of its large 6mm ablator shell diameter) at these densities while using NIF's current energy limitations.
Instead, Revolver will utilize either a 2 photon-process printed lattice at $\sim5$ mg/cc or a tent to support the driver shell inside the ablator. Simulations have demonstrated that a sufficiently low density foam or a tent will have minimal effects on the ablator implosion energetics. Therefore, the sub-scale OMEGA targets were designed with vacuum between the shells, with the inner Cr shell being centered in the Be ablator via a gold cone to simplify target fabrication; See Fig.~\ref{radiograph}A. During the implosion, the inner shell material and cushion layer (described in the following paragraph) expand into the vacuum between shells due to X-ray preheat and are subsequently recompressed during the collision. The primary goal of these experiments is to validate our computational model of the collision process in such a configuration, including the ability to predict the energy transfer between the shells.

\begin{figure}
\includegraphics[width=.5\textwidth]{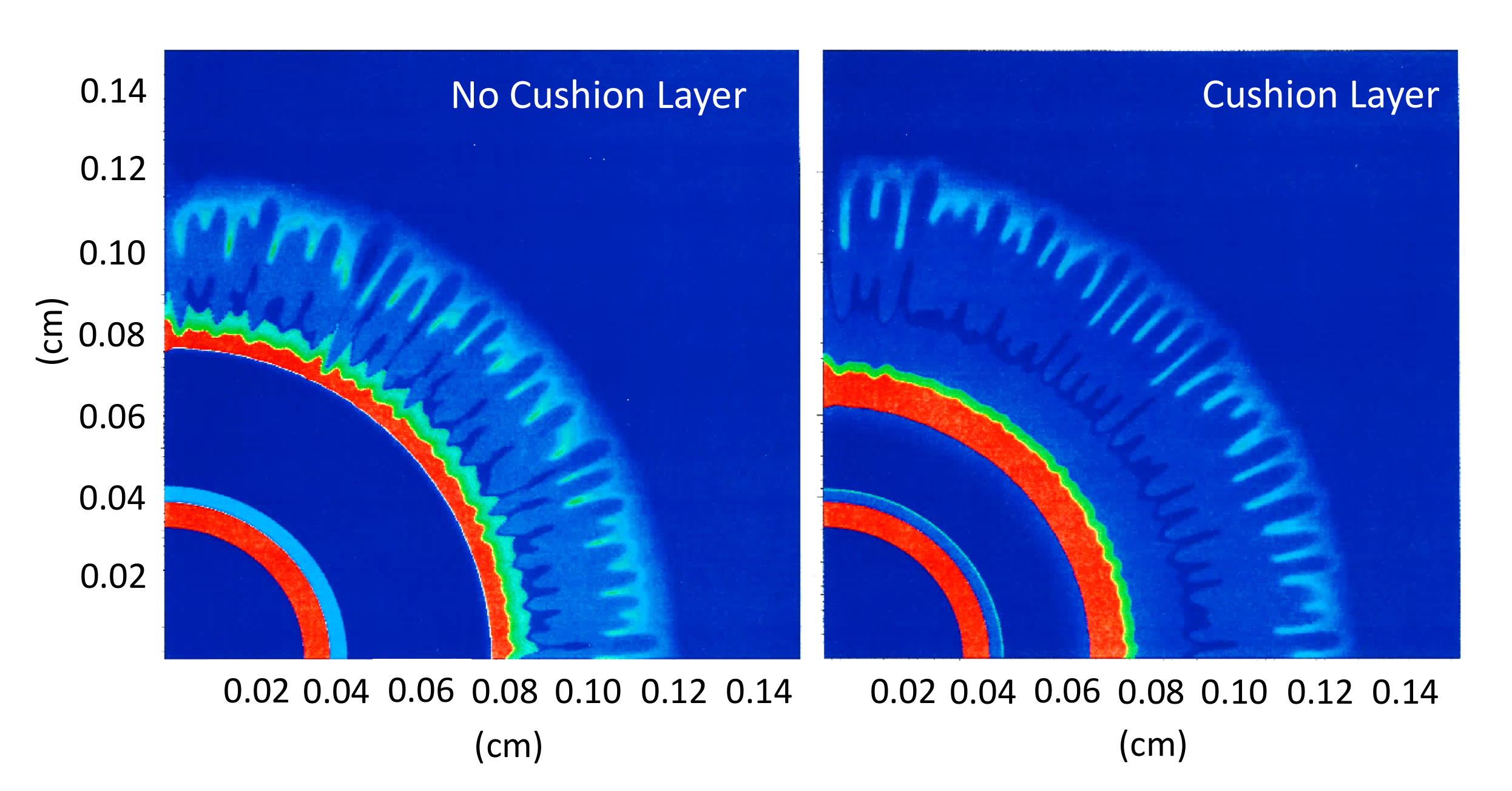}
\caption{\label{cushion}Simulations of an ignition scale Revolver capsule showing the mode-smoothing effect of a cushion layer on the middle driver shell (Right) in comparison with a case with no cushion layer (Left).  The initial radius of the middle shell is 0.115 cm.}
\end{figure}

The inner Cr shell was coated with a 23 $\mu$m thick parylene cushion layer. The cushion layer is similar in placement to the tamping layer in double shell capsule designs \cite{2018PhPl...25i2706M}. However, the double shell tamper has the express purpose of tamping the inner tungsten shell from expansion due to x-ray preheat while lowering the Atwood number between the foam and inner shell. For Revolver targets, the cushion layer serves as a low pass filter for high frequency spatial modes on the ablator induced by laser intensity non-uniformities and amplified through the acceleration phase via the Rayleigh Taylor instability on the outer surface and deceleration on the inner surface. During the shell collision, the spikes and bubbles from the Rayleigh-Taylor growth result in high Legendre mode pressure variations which, in the absence of a cushion layer, results in the direct imprint of these modes onto the inner shell. The cushion layer provides a layer of material in which these pressure variations can partially isotropize during the collision. In the absence of this layer, the isotopization occurs in the outer layer of the driver shell and induces mass modulations. A characteristic length scale over which this smoothing takes place can be estimates as the distance a sound wave can transit during the collision
\begin{equation}
\Delta x_{smooth}\sim c_{s, cush.}\Delta t_{coll.},
\end{equation}
where $c_{s, cush.}$ is the sound speed in the cushion and $\Delta t_{coll.}$ is the collision time which is approximately the width of the pressure pulse in the cushion material. The importance of the cushion layer is clearly demonstrated in simulations of the ignition design shown in Fig.~\ref{cushion}. Without the cushion, the middle driver shell has significant modulations after its collision with the ablator. Thus, understanding the energy requirements and smoothing capabilities of cushion layers is crucial for optimizing directly-driven multi-shell ignition designs 

\subsection{Experiment Setup}
The target, with cone axis aligned to TIM 5 in the OMEGA target chamber, was driven by 39 of OMEGA's 60 beams pointed at target chamber center (TCC) using a 1-ns square pulse. The average beam power was 223J/beam, resulting in 8.7 kJ delivered to the target. The beams on target utilized SG-5 phase plates and the resulting average irradiance on the target was $2.9\times10^{14} \ \textrm{W} \ \textrm{cm}^{-2}$, similar to the irradiance in the Revolver ignition scale design. The positioning of the target cone towards TIM 5 also facilitated the exploratory use of the streaked optical pyrometer (SOP) and velocity interferometer system for any reflector (VISAR)\cite{2007RScI...78c4903M} or high-speed photon doppler velocimetry \cite{2017APS..SHK.O2005M} diagnostics to measure the interior of the inner shell. Development of these techniques in the present target design is ongoing.

The primary diagnostic for these shots were two Fe backlighter foils which produce He-$\alpha$ emission at 6.67 keV. Synthetic radiograph images of simulated implosions produced a clear sharp edge at these energies, which corresponds to the outer edge of the chromium shell. The backlighters, when combined with a 6$\mu m$ Fe camera filter to reduce stray Bremsstrahlung emission, produced similarly sharp images (See Fig.~\ref{radiograph}B). Each backlighter was driven with 4 to 7 beams which were defocused to produce an illuminated area with an irradiance of $7.5\times10^{14} \ \textrm{W} \ \textrm{cm}^{-2}$. Images for each backlighter were obtained via a 4 by 4 array of $15\mu m$ pinholes with a magnification of 4 at the image plane. The images were recorded on CCDs of Xray Framing Camera 1 and Sydor Framing Camera 4 placed in TIMs 2 and 4, respectively. Each pixel on the CCD corresponded to a physical distance of $4.5\mu m$ at the target. The calibrated timing of the 4 strips of the backlighter cameras were determined offline prior to the shot. In addition to the backlighter, self emission imaging was also utilized to determine the ablator trajectory in the first $\sim 1.3 ns$ of the implosion. The self emission images were recorded on Sydor Framing Camera 3 in TIM 1 using a similar pinhole array to the backlighter, but with a $12 \mu m$ thick Be filter.

\section{Experiment Results and Simulations} 

Backlit and self emission images of the Cr and Be shells provided data for comparison with simulated trajectories of the ablator and driver shells. These images span 5 different shots with identical drive energy and pulse shape requests. Self emission images of the ablator were processed to determine the radius of the maximum gradient on the inside of the image intensity maximum by using a Sobel edge detection algorithm in Matlab\cite{MATLAB:2017b}. A circle was fit to a portion of the edge away from the cone using the method outlined in Appendix A. The radius of the inner shell was determined using a similar method, also outlined in Appendix A. Due to the high opacity of the inner shell at the backlighter wavelength, the radius determined by the edge detection method corresponds to the outer edge of the chromium shell. A comparison of a pre-shot radiograph and backlit image of the imploding Cr shell is shown in Fig.~\ref{radiograph}.

\begin{figure}
\includegraphics[width=.5\textwidth]{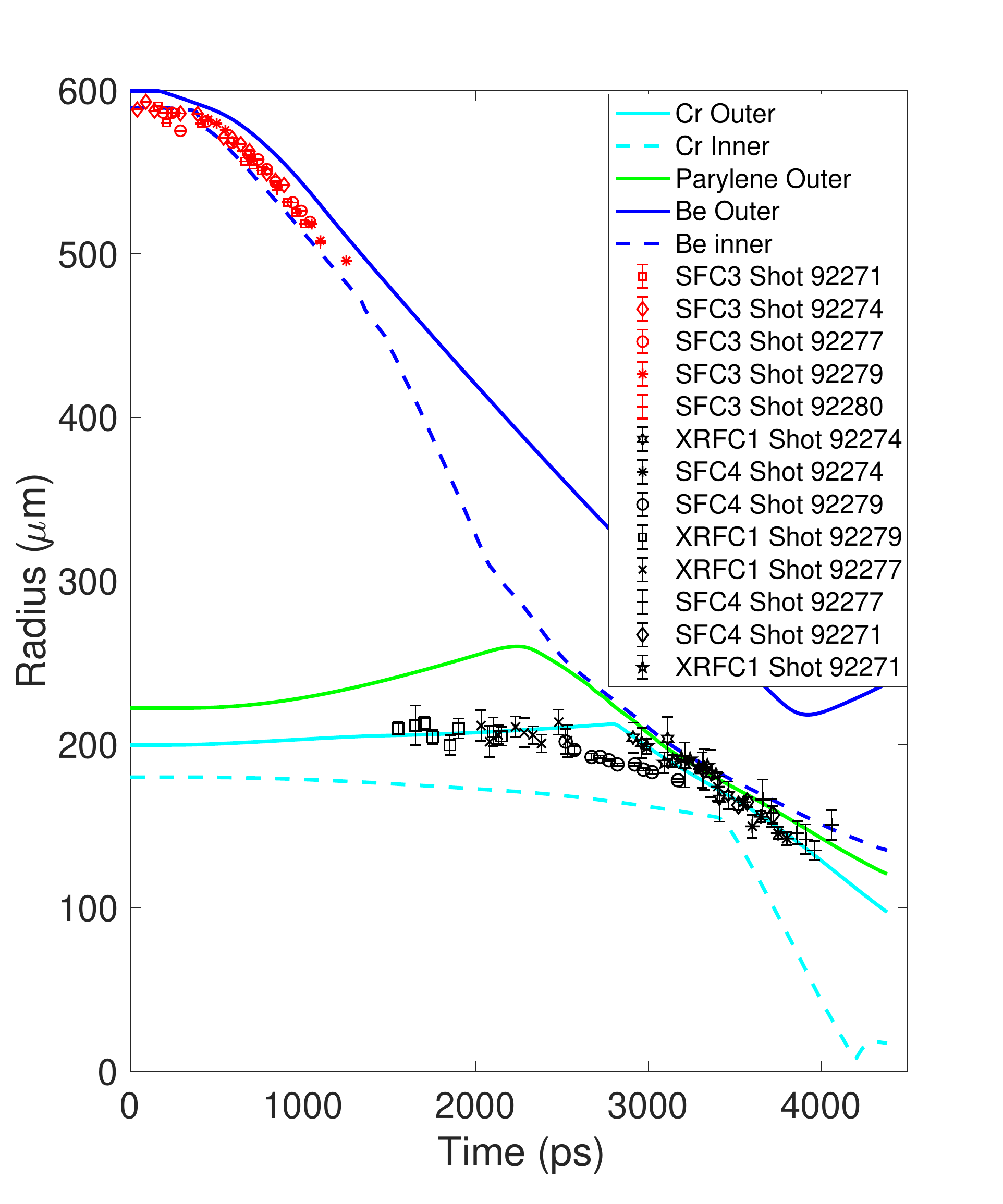}
\caption{\label{sim} A comparison of the measured ablator (red markers) and driver (black markers) shell radii with simulation data. The dotted (solid) lines represent the inner (outer) edges of the two shells and the parylene coating (green) on the outside of the inner shell.}
\end{figure}

To validate the collision model, post shot simulations using characteristic as-shot quantities were conducted using the the radiation hydrodynamics code HYDRA\cite{1996PhPl....3.2070M} and are summarized in Table~\ref{values}. For simplicity, the experiments are compared with simulations of a symmetric implosion, neglecting the effect of the cone. Since the measured shell radius was determined using an arc of the shell which was out of hydrodynamic communication with the cone, the difference is expected to be negligible. The measured radii from the self emission and backlit images are compared to post-shot simulation results in Fig.~\ref{sim}. The measured values of the shell positions are in excellent agreement with simulations, with the exception of shot 92279 due to an offset of the inner shell centering incurred during assembly. In particular, a linear fit to the measured outer Cr surface between 3.1 and 3.8 ns resulted in a velocity of $7.52\pm0.59$ cm/$\mu$s, which is in excellent agreement with the simulated value of 7.27 cm/$\mu$s. 

\begin{table}
\begin{tabular}{lllll}
 Outer Be Radius&   606$\mu$m\\
 Inner Be Radius& 589.5$\mu$m  \\
  \% Ar in Be& 0.8\% Shell \\
 Outer Cr Radius & 200$\mu$m \\
 Inner Cr Radius & 180$\mu$m\\ 
 Parylene thickness & 23$\mu$m\\
 Laser Energy & 223.1J/beam

\end{tabular}
\caption{\label{values}Parameters used for the post-shot simulations.}
\end{table}

\begin{figure}
\includegraphics[width=.5\textwidth]{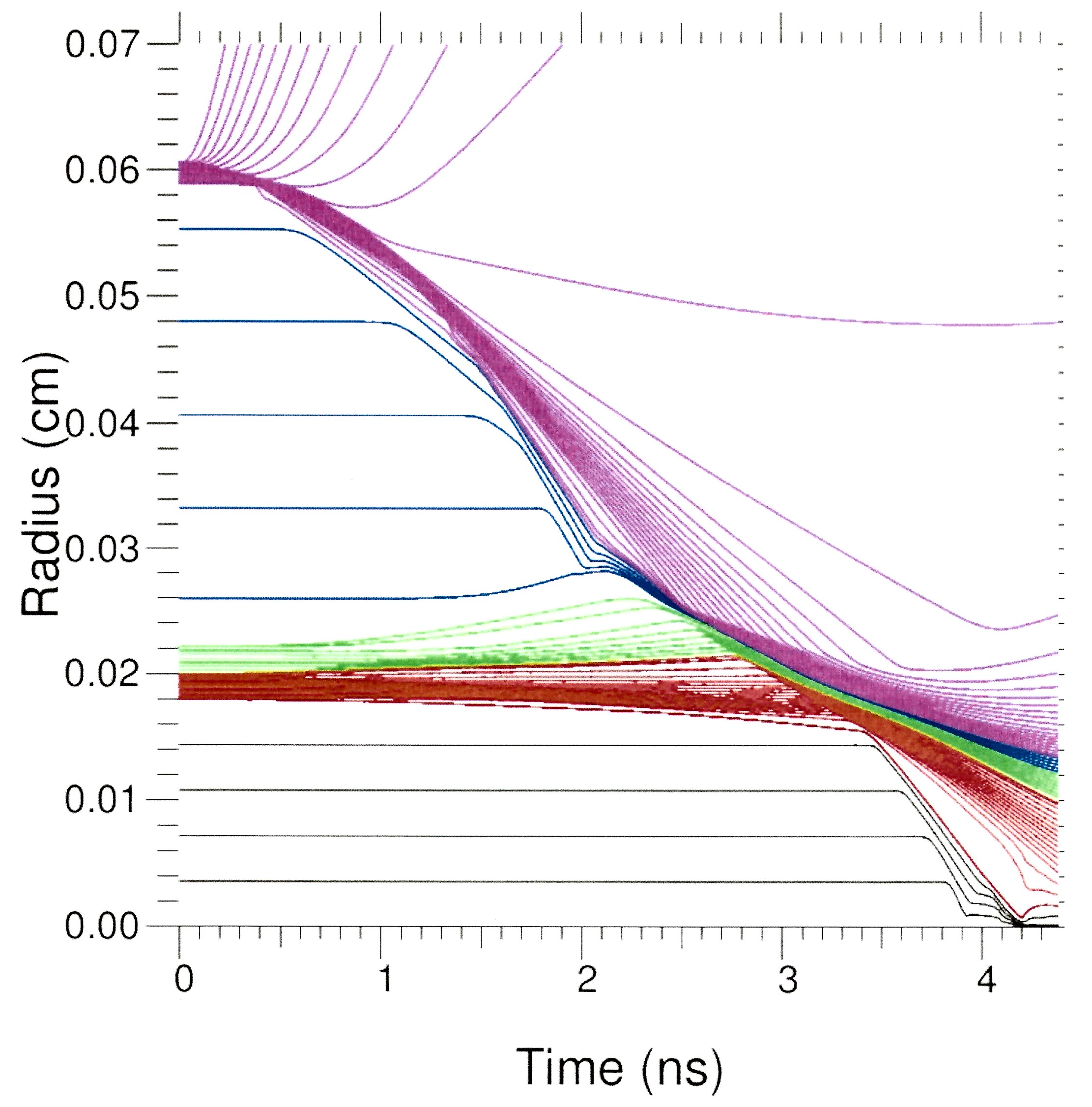}
\caption{\label{ddd} A Lagrangian plot from a simulation of the OMEGA experiment. The colors represent different materials with magenta, green, and red being Be, Parylene, and Cr, respectively. All other colors represent vacuum.  }
\end{figure}

The agreement of the measured outer Cr surface trajectory with the simulated values suggest that the energy transfer between shells in the experiments were similar to that in the simulation. For the simulated values, the symmetric 60 beam equivalent kinetic energy in the ablator payload mass is 1.16kJ and the payload kinetic energy of the inner shell is 0.42 kJ, suggesting an energy transfer of 36\%. Due to the close agreement between simulations and experiments, we infer a similar efficiency for these experiments. Extracting a measured efficiency from experiments is not possible in these sub-scale experiments due to the velocity dispersion of the inner shell, as can be seen at times greater than 3.5 ns in Fig.~\ref{ddd}. Such effects are not an issue in the ignition scale design (compare Fig.~1) where the driver shell payload mass is well confined. 

While similar energy transfer efficiencies have been observed on recent indirect drive double shell experiments on the NIF, an efficiency of 36\% is much lower than the design value of 59\% at full scale. This difference was expected for these experiments and the reason for the difference can be traced to their sub-scale nature. The amplification of ablator pressure during implosion is optimally proportional to the square of the convergence since $\rho\propto C^2$ when the shell thickness is constant. In these experiments, the ablator shell is not well confined due to limitations on the pulse length set by the diagnostic needs of the experiment. The ignition design ablator is compressed by three shocks during the drive while only one exists in the present experiments. The result is a greater conversion of shell kinetic energy into ablator internal energy in the recompression of the ablator during the shell collision, see Fig.~\ref{ddd}. Likewise, a reduction in ablator pressure results in a lower drive pressure imparted by the ablator on the driver shell. As the parylene layer compresses during the collision, a shock is driven through the parylene into the Cr. 
After this shock breaks out of the inside surface of the Cr shell, a rarefaction wave propagates back into the Cr shell.
As a result, the inner surface of the Cr shell moves with a velocity greater than the outer surface. In the ignition scale design, the increased pressure during the shell collision results in a second shock that improves the spatial confinement of the shell.

\section{Conclusion}

These experiments have demonstrated the predictive capability of simulations modeling the shell collision process for the Revolver target design. The design of the outer two shells in Revolver targets are unique in that they minimizes the amount of material between the ablator and driver shells. In the absence of $\sim40-80$ mg/cc foam, a cushion layer on the inner shell partially blows off due to penetrating X-ray flux from the corona. This layer is recompressed during the collision and together with the unablated cushion material serves to mediate the collision and filter high-Legendre-mode growth that would otherwise imprint on the driver shell. These experiments demonstrate the ability of simulations to correctly model the shell collision process and further demonstrates that there are no significant unforeseen energy loss mechanisms, such as those stemming from insufficient computational models, during the collision. 

The experiments also demonstrate for the first time the measurement of the post-collision trajectory for a high density inner shell. To date, measured collisions of double shell targets have used low density surrogate inner shells made of SiO$_2$ (2.2 g/cc) to facilitate radiography. The low density inner shells mean that previously measured collisions with Al (2.7 g/cc) and CH (1.04 g/cc) ablators provided Atwood numbers different from the full scale experiments which they were meant to emulate. The present experiments provide a measurement of shell collisions with higher Atwood numbers than previous experiments.    

These experiments provide confidence in our ability to model shell collisions having parameters which are relevant to the Revolver ignition design. However, full scale experiments at the NIF will be needed to fully assess the shell collision process at longer time scales and higher pressures encountered in these designs. 

\section*{Acknowledgments }
The authors would like to thank Charles Sorce, James Tellinghuisen, Ray Bahr, Timothy Filkins, and the support staff at LLE who assisted during the shot day. This work was supported by the Laboratory Directed Research and Development program of Los Alamos National Laboratory under project number 20180051DR.

\section*{Appendix: Processing of Shot Data}

Backlit images such as the one in Fig.~\ref{radiograph} are processed by detecting edges which correspond to large gradients in the image. The edge of interest is the boundary between the opaque chromium shell and the transparent surrounding material. The length of a chord a distance h inward from the edge of a circle of radius R is $L=2\sqrt{h(2R-h)}$. Taking h to be the physical size of the target projected onto a pixel results in $L\approx88\mu m$, while for $h \approx 13\mu m$, the pinhole resolution, $L\approx141\mu m$. Both of these path-lengths L result in near complete attenuation of 6.67 keV X-rays, therefore the detected edge at the outer surface is an excellent approximation to the target outer surface radius. Once the edges are obtained, the one corresponding to the chromium shell is selected for analysis.

The radius is determined by performing a least squares fit of a circle to an arc of the target surface opposite to the direction of the cone. Depending on the image contrast, a section corresponding to an angular extent in the range $90^o<\theta<150^o$ is selected, avoiding the region from the target equator to the intersection of the cone with the shell. The method used to fit the data is that of Ref.~\onlinecite{Gander1994}, an expansion of which is given in this appendix.

A circle is fit to the selected data by considering the algebraic representation of a circle in the 2D plane with vectors $\vc{x},\vc{b}\in\mathbb{R}^2$ such that
\begin{equation}\label{circ}
a\vc{x}^T\vc{x}+\vc{b}^T\vc{x}+c=0.
\end{equation}
Inserting the coordinates of the N points on the selected arc into Eq.~\ref{circ} gives the overdetermined linear system $B\vc{u}=0$ where $\vc{u}=(a,b_1,b_2,c)^T$ and
\begin{gather}
B=
  \begin{bmatrix}
    x_{11}^2+x_{12}^2 & x_{11} & x_{12}& 1  \\
    \vdots&\vdots&\vdots&\vdots \\
    x_{N1}^2+x_{N2}^2 & x_{N1} & x_{N2}& 1  
  \end{bmatrix}.
\end{gather}
When a=1, the system can be written in the form $A\vc{v}=\vc{f}$
\begin{gather}
  \begin{bmatrix}
    x_{11} & x_{12}& 1  \\
    \vdots&\vdots&\vdots\\
     x_{N1} & x_{N2}& 1  
  \end{bmatrix}
    \begin{bmatrix}
	b_1\\
	b_2\\
	c
  \end{bmatrix}=
    \begin{bmatrix}
    -(x_{11}^2+x_{12}^2) \\
    \vdots\\
    -(x_{N1}^2+x_{N2}^2 ) 
  \end{bmatrix}.
\end{gather}
To solve for the vector $\vc{v}=(b_1,b_2,c)^T$ in the overdetermined system, the method of least squares is used, the solution being 
\begin{equation}
\vc{v}=(A^TA)^{-1}A^T\vc{f}.
\end{equation}
The solution for $\vc{v}$ provides the center and radius of the circle as 
\begin{equation}
\vc{x}_c=\big(-\frac{b_1}{2},-\frac{b_2}{2}\big)
\end{equation}
and
\begin{equation}
R=\sqrt{\frac{b_1^2+b_2^2}{4}-c}.
\end{equation}
Error bars in the the radius R are calculated by propagation of error for the errors of each parameter $b_1,\ b_2,\ c$ calculated from the covariance matrix. It is assumed that errors in the parameters are uncorrelated.

\bibliography{draftv-2}

\end{document}